\documentclass[preprint,prc,aps,epsfig]{revtex4}
\usepackage{psfig}
\def\beq{\begin{equation}}
\def\eeq{\end{equation}}
\def\beqa{\begin{eqnarray}}
\def\eeqa{\end{eqnarray}}
\def\MeV{\nobreak\,\mbox{MeV}}
\def\GeV{\nobreak\,\mbox{GeV}}

\def\Tr{\mbox{ Tr }}

\def\me#1{\langle{#1}\rangle}

\def\bra#1{\langle #1|}
\def\ket#1{| #1\rangle}
\def\qbar{\overline{q}}
\def\gs{g_{\rm s}}
\def\G{{\cal G}}
\def\mixbar{\gs\qbar\sigma\!\cdot\!\G q}
\def\mixs{\gs\bar{s}\sigma\!\cdot\!\G s}

\def\gluoncon{{\displaystyle{\gs^2G^2}}}
\def\qslash{\rlap{/}{q}}
\def\xsla{\rlap{/}{x}}
\def\qsq{q^2}

\begin{document}

\title{\sc  Are $\Theta^+$ and the Roper resonance diquark-diquark-antiquark
states?}
\author{R.D. Matheus, F.S. Navarra, M. Nielsen, R. Rodrigues da Silva}
\affiliation{Instituto de F\'{\i}sica, Universidade de S\~{a}o Paulo\\
 C.P. 66318,  05315-970 S\~{a}o Paulo, SP, Brazil}
\author{and S.H. Lee}
\affiliation{Institute of Physics and Applied Physics, Yonsei University,
Seoul 120-749, Korea}

\begin{abstract}
We consider  a $[ud]^2\bar{s}$ current in the QCD sum rule
framework to study the mass of the recently observed pentaquark state 
$\Theta^+(1540)$, obtaining  good agreement with the experimental value. 
We also study the mass of the pentaquark
$[ud]^2\bar{d}$. Our results are compatible with the interpretation of the 
$[ud]^2\bar{d}$ state as being the Roper resonance $N(1440)$, as suggested 
by Jaffe and Wilczek.
\end{abstract}

\maketitle

\vspace{1cm}

The recent discovery of an exotic baryon with the $K^+ n$ quantum numbers, 
the $\Theta^+ (1540)$ \cite{lep,diana,clas,saphir}
with mass and width compatible with predictions
made by Diakonov, Petrov and Polyakov \cite{dpp} prompted a lively debate 
about the spectroscopy of non conventional hadronic states 
\cite{jawil,kali,hep07,zhu}. Since the 
$\Theta^+$ cannot be a three quark state and its minimal quark content is 
$u u d d \overline{s}$ one is left with the question of how these quarks are 
organized.  They could be: a) uncorrelated quarks inside a bag \cite{strot}; 
b) a $K-N$ molecule bound by a van-der Waals force \cite{brodsky}; c) a 
``$K-N$'' bound state in which $u u d$ and $u \overline{s}$ are not separately 
in color singlet states \cite{zhu}; d) a diquark-triquark 
$(u d) - (u d \overline{s})$ 
bound state \cite{kali} and 
e) a diquark-diquark-antiquark state \cite{jawil}. In this note we will 
explore this last possibility, using the QCD sum rules framework to give a
more quantitative basis to the semi-qualitative argument presented in 
\cite{jawil}. 

According to Jaffe and Wilczek (JW), each diquark pair 
has spin zero and is in the $\bar{\textbf{3}}$ representation of SU(3), 
in color and flavor. The two diquark pairs combine in a $P$-wave orbital 
angular momentum
to form a  $\textbf{3}$ state in color, 
spin $S=0$, and $\bar{\textbf{6}}$ in flavor.  
The resulting state is then combined with
the antiquark to form a flavor antidecuplet $\bar{\textbf{10}}_f$
and octet ${\textbf{8}_f}$, with spin $S=1/2$.
The $\Theta^+$ is at the top of the antidecuplet $\bar{\textbf{10}}_f$
and has an isospin $I=0$. JW have also interpreted the lightest particle 
in the 
octet ${\textbf{8}_f}$, $[ud]^2\bar{d}$, as the Roper resonance, since it 
has the same quantum 
numbers of the nucleon.  The Roper  resonance is thus
identical to the $\Theta^+$ except for the 
substitution of the strange antiquark by a down antiquark. This would  
explain why the mass difference between 
$\Theta^+(1540)$ and  $N(1440)$ is so close to the strange quark mass.
It is therefore interesting to verify if this mass splitting can be obtained 
in a more detailed calculation. In the case of QCD sum rules (QCDSR) there
are several contributions from the OPE, involving many operators which 
account for the nonperturbative dynamics, and thus  this simple dependence 
with $m_s$ is not expected a priori. Using the same 
diquark-diquark-antiquark for the Roper, we will calculate this mass 
difference.
 
Among the works on the $\Theta^+$ resonance 
the QCDSR calculation of Zhu \cite{zhu} is of 
particular relevance for us. He used the triquark - $q \overline{q}$ 
configuration with both in the  color adjoint representation. In fact, 
this configuration is the c) in the short  list given above. In QCDSR, 
different configurations are implemented with the use of different 
interpolating currents. As it will be seen, both currents will give 
approximately the same mass for the $\Theta^+$ but not necessarily the 
same mass splitting between $\Theta^+$ and Roper resonance.


In the QCDSR approach \cite{svz,rry}, the short range perturbative QCD is
extended by an OPE expansion of the correlators, which results in 
a series in powers of
the squared momentum with Wilson coefficients. The convergence at low
momentum is improved by using a Borel transform. The expansion involves
universal quark and gluon condensates. The quark-based calculation of
a given correlator is equated to the same correlator, calculated using
hadronic degrees of freedom via a dispersion relation, providing sum rules
from which a hadronic quantity can be estimated. The QCDSR 
calculation of hadronic masses centers around the two-point
correlation function given by
\beq
\Pi(q)\equiv i\int d^4 x\, e^{iq\cdot x}
\bra{0} T\eta(x)\overline{\eta}(0)\ket{0}\ ,
\label{cor}
\eeq
where $\eta(x)$ is an interpolating field (a current) with the quantum
numbers of the hadron we want to study.

Following the JW conjecture \cite{jawil}, we can write two 
independent interpolating fields with the quantum numbers of $\Theta^+(1540)$:
\beqa
\eta_1(x)&=&{1\over\sqrt{2}}\epsilon_{abc}\left([u_a^T(x) C \gamma_5 d_b(x)]
[u_c^T(x) C 
\gamma_5 d_e(x)]C\bar{s}^T_e(x) - (u\leftrightarrow d)\right)\ ,
\label{eta1}
\\*
\eta_2(x)&=&{1\over\sqrt{2}}\epsilon_{abc}\left([u_a^T(x) C  d_b(x)]
[u_c^T(x) C  d_e(x)]C\bar{s}^T_e(x) - (u\leftrightarrow d)\right)\ ,
\label{eta2}
\eeqa
where $a,~b,~c$ and $e$ are color index and $C=-C^T$ is the charge 
conjugation operator. One can check that each diquark
pair has spin zero and is in the $\bar{\textbf{3}}$ representation of SU(3) 
in color and flavor. The total current has isospin zero, positive parity 
and spin 1/2. 

As in the nucleon case, where one also has two independent currents with 
the nucleon quantum numbers \cite{io1,dosch}, the most general current for
$\Theta^+$ is a linear combination of the currents given above:
\beq
\eta(x)=\left[t\eta_1(x) + \eta_2(x)\right] ,
\label{cur}
\eeq
with $t$ being an arbitrary parameter. In the case of the nucleon, the 
interpolating field with $t=-1$ is known as Ioffe's current \cite{io1}.
This current maximizes the overlap with the nucleon as compared with
the excited states, and minimizes the contribution of higher
dimension condensates.

Inserting Eq.~(\ref{cur}) into Eq.~(\ref{cor}) we obtain
\beq
\bra{0} T\eta(x)\overline{\eta}(0)\ket{0}=
t^2\Pi_{11}(x)+2t\left(\Pi_{12}(x)+\Pi_{21}(x)\right)+\Pi_{22}(x)\,
\eeq
Calling $\Gamma_1=\gamma_5$ and $\Gamma_2=1$ we get
\beqa
\Pi_{ij}(x)=\bra{0} T\eta_i(x)\overline{\eta}_j(0)\ket{0}=
\epsilon_{abc}\epsilon_{a'b'c'}C{S_{e'e}^s}^T(-x)C\left\{
-\Tr[\Gamma_iS_{bb'}(x)\Gamma_jCS^T_{aa'}(x)C]
\right.
\nonumber\\
\times\Tr[\Gamma_iS_{ee'}(x)\Gamma_jCS^T_{cc'}(x)C]
+\Tr[\Gamma_iS_{be'}(x)\Gamma_jCS^T_{cc'}(x)C\Gamma_iS_{eb'}(x)\Gamma_j
CS^T_{aa'}(x)C]
\nonumber\\
+\Tr[\Gamma_iS_{bb'}(x)\Gamma_jCS^T_{ca'}(x)C\Gamma_iS_{ee'}(x)\Gamma_j
CS^T_{ac'}(x)C]-\Tr[\Gamma_iS_{be'}(x)\Gamma_jCS^T_{ac'}(x)C]
\nonumber\\
\times\Tr[\Gamma_iS_{eb'}(x)\Gamma_jCS^T_{ca'}(x)C]
-\Tr[\Gamma_iS_{ab'}(x)\Gamma_jCS^T_{ea'}(x)C\Gamma_iS_{ce'}(x)\Gamma_j
CS^T_{bc'}(x)C]
\nonumber\\
-\Tr[\Gamma_iS_{ba'}(x)\Gamma_jCS^T_{cb'}(x)C\Gamma_iS_{ec'}(x)\Gamma_j
CS^T_{ae'}(x)C]+\Tr[\Gamma_iS_{ba'}(x)\Gamma_jCS^T_{ab'}(x)C]
\nonumber\\
\left.
\times\Tr[\Gamma_iS_{ce'}(x)\Gamma_jCS^T_{ec'}(x)C]
+\Tr[\Gamma_iS_{bc'}(x)\Gamma_jCS^T_{ae'}(x)C]\Tr[\Gamma_iS_{ea'}(x)\Gamma_j
CS^T_{cb'}(x)C]\right\}\;,
\label{expa}
\eeqa
where $S_{ab}(x)$ and $S_{ab}^s(x)$ are the light and strange quark
propagators respectively.

In order to evaluate the correlation function $\Pi(q)$ at the quark
level, we first need to determine the quark propagator in the
presence of quark and gluon condensates. Keeping track of the terms
linear in the quark mass and taking into account quark and gluon 
condensates, we get \cite{yhhk}
\beqa
S_{ab}(x)&=&\bra{0} T[q_a(x)\overline{q}_b(0)]\ket{0}={i\delta_{ab}\over2
\pi^2x^4}\xsla-{m_q\delta_{ab}\over4\pi^2x^2}-{i\over32\pi^2x^2}t^A_{ab}
\gs G^A_{\mu\nu}
(\xsla\sigma^{\mu\nu}+\sigma^{\mu\nu}\xsla)
\nonumber\\
&-&{\delta_{ab}\over12}\me{\qbar q}
-{m_q\over32\pi^2}t^A_{ab}\gs G^A_{\mu\nu}\sigma^{\mu\nu}\ln(-x^2)
+{i\delta_{ab}\over48}m_q\me{\qbar q}\xsla-{x^2\delta_{ab}
\over2^6\times3}\me{\mixbar}
\nonumber\\
&+&{ix^2\delta_{ab}\over2^7\times3^2}m_q\me{\mixbar}\xsla
-{x^4\delta_{ab}\over2^{10}\times3^3}\me{\qbar q}\me{\gluoncon}\,,
\eeqa
where we have used the factorization approximation for the multi-quark
condensates, and we have used
the fixed-point gauge \cite{yhhk}.

Lorentz covariance, parity and time reversal imply that the two-point
correlation function in Eq.~(\ref{cor}) has the form
\beq
\Pi(q)= \Pi_1(q^2)+ \Pi_q(\qsq) \qslash\;.
\label{stru}
\eeq
A sum rule for each scalar invariant function $\Pi_1$ and $\Pi_q$, can be 
obtained. As in ref.~\cite{zhu}, in this work we focus on the chirality
even structure $\Pi_q(\qsq)$.

The phenomenological side is described, as usual, as a 
sum of pole and continuum, the latter being approximated  by the OPE 
spectral density.
In order to suppress the condensates of higher dimension and at the same time
reduce the influence of higher resonances  we perform a  
standard Borel transform \cite{svz}: 
\beq
\Pi (M^2) \equiv \lim_{n,Q^2 \rightarrow \infty} \frac{1}{n!} (Q^2)^{n+1} 
\left( - \frac{d}{d Q^2} \right)^n \Pi (Q^2)
\eeq
($Q^2 = - q^2$) with the squared Borel mass scale $M^2 = Q^2/n$ kept 
fixed in the limit.

After Borel transforming each side of $\Pi_q(Q^2)$ and transferring
the continuum contribution to the OPE side we obtain the following
sum rule at order $m_s$:
\beqa 
\lambda_{\Theta}^2 e^{-m_\Theta^2/M^2} &=&{3c_1\over 2^{11}\times7!~
\pi^8}M^{12}
E_5+{m_s\me{\bar{s}s}c_1\over2^{10}\times5!~\pi^6}M^{8}E_3+
{\me{\gluoncon}c_2\over2^{15}\times5!~\pi^8}M^{8}E_3
\nonumber\\
& + &
{\me{\qbar  q}^2c_3\over2^{9}\times3^2~\pi^4}M^{6}E_2-
{m_s \me{\mixs}c_1\over2^{14}\times3^2~\pi^6}M^{6}E_2
\nonumber\\
& + &{m_s \me{\bar{s}s}\me{\qbar q}^2c_3\over2^{6}\times3^2~\pi^2}M^{2}E_0
+{\me{\qbar q}^4c_1\over6^3}\,,
\label{sumq}
\eeqa
where $c_1 = 5t^2+2t+5$, $c_2 = (1-t)^2$, $c_3=7t^2-2t-5$ and
we have defined
\beq
E_n\equiv 1-e^{-s_0/M^2}\sum_{k=0}^n\left(s_0\over M^2\right)^k{1\over k!}
\ ,
\label{con}
\eeq
which accounts for the continuum contribution with $s_0$ being the continuum
threshold \cite{io1}.

To extract the $\Theta^+$ mass, $m_\Theta$, we take the derivative of 
Eq.~(\ref{sumq}) with respect to $M^{-2}$ and divide it by Eq.~(\ref{sumq}).

The interpolating field  for $N^+$ is also given by Eq.~(\ref{cur}), just by
changing $\bar{s}$ by $\bar{d}$ in Eqs.~(\ref{eta1}) and (\ref{eta2}).
Therefore, the sum rule for $N^+$ can be obtained from the sum
rule in Eq.~(\ref{sumq}) by neglecting the terms proportional to $m_s$.


In the complete theory,  the mass extracted from the sum rule
should be independent of the Borel mass $M^2$. However, in  a truncated 
treatment there will 
always  be some dependence left.  Therefore, one has to work in a region 
where the approximations made are supposedly acceptable and where 
the result depends only moderately on the Borel variables. 

In the numerical analysis of the sum rules, the values used for the strange
quark mass and condensates are: $m_s=0.15\,\GeV$, $\me{\qbar q}=\,
-(0.23)^3\,\GeV^3$,
$\langle\overline{s}s\rangle\,=0.8\me{\qbar q}$,  $\me{\mixs}=m_0^2
\me{\bar{s}s}$ with $m_0^2=0.8\,\GeV^2$ and $\me{\gluoncon}=0.5~\GeV^4$.

\begin{figure} \label{fig1}
\centerline{\psfig{figure=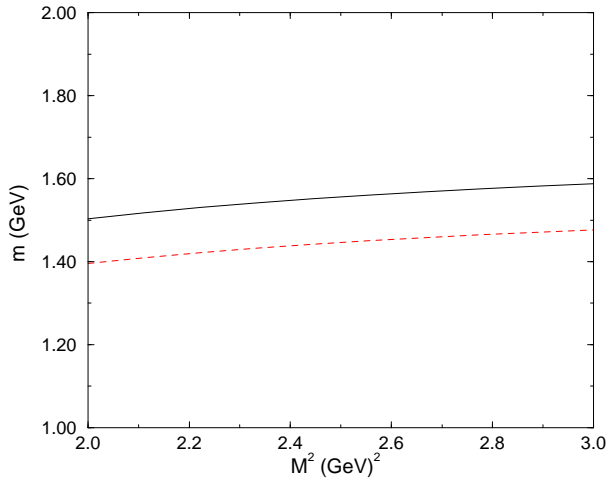,width=8cm,angle=0}}
\caption{The masses of the pentaquarks $\Theta^+$ (solid line) and
$N^+$  (dashed line)as a function of the Borel parameter $M^2$.}
\end{figure}

We evaluate our sum rules in 
the range $2.0\leq M^2\leq3.0\GeV^2$. In Fig.~1 we show the masses
of the pentaquarks $\Theta^+$ and $N^+$, as a function of the Borel mass
using $s_{0\Theta}=(1.5+0.5)^2~\GeV^2=4.0~\GeV^2$, 
$s_{0N}=(1.4+0.5)^2~\GeV^2=3.6~\GeV^2$ and the current parameter in 
Eq.~(\ref{cur}), $t=-1$. We can see that the results are reasonably stable
as a function of the Borel mass in the considered interval, and that
the values obtained for the masses are in agreement with the experimental
values $m_{\Theta}=1540~\MeV$ and $m_{\rm{Roper}}=1440~\MeV$. Therefore,
from our results, it is really possible that the Roper resonance can 
be identified with the pentaquark $[ud]^2\bar{d}$, as suggested by
Jaffe and Wilczek \cite{jawil}. It is very important to mention that
the only free parameters in our calculations are the continuum thresholds
and the current parameter, and that we have just used the most conventional
values for these parameters : $s_0=(m+0.5~\GeV)^2$ and $t=-1$. Therefore,
the fact that the masses obtained for both cases are in agreement with 
the experimental results can not be underestimated.

\begin{figure}[h] \label{fig2}
\centerline{\psfig{figure=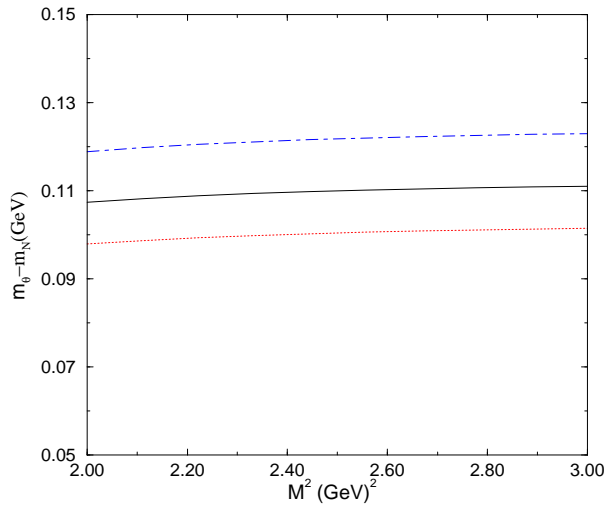,width=8cm,angle=0}}
\caption{$s_0$ dependence of the mass difference. The solid, dot-dashed and
dotted lines give $m_\Theta-m_N$ for $s_{0\Theta}=4.0~\GeV^2$ and 
$s_{0N}=3.6~\GeV^2$, $s_{0\Theta}=3.6~\GeV^2$ and $s_{0N}=3.2~\GeV^2$,
and $s_{0\Theta}=4.4~\GeV^2$ and $s_{0N}=4.0~\GeV^2$ respectively.}
\end{figure}

In Fig.~2 we show the mass difference, $m_\Theta-m_N$, as a function
of the Borel mass, for different values of the continuum thresholds
but keeping $t=-1$.
We consider continuum thresholds in the ranges $3.6~\GeV^2\leq s_{0\Theta}\leq
4.4~\GeV^2$ and $3.2~\GeV^2\leq s_{0N}\leq4.0~\GeV^2$.
The effect of increasing (decreasing) the continuum
threshold is to increase (decrease) the masses, however, the 
increase (decrease) in the $N^+$ mass is bigger than in the $\Theta^+$ mass,
leading to the opposite behavior in the mass difference. For all values
considered we can see that the mass difference is of order of $100~\MeV$.

\begin{figure}[h] \label{fig3}
\centerline{\psfig{figure=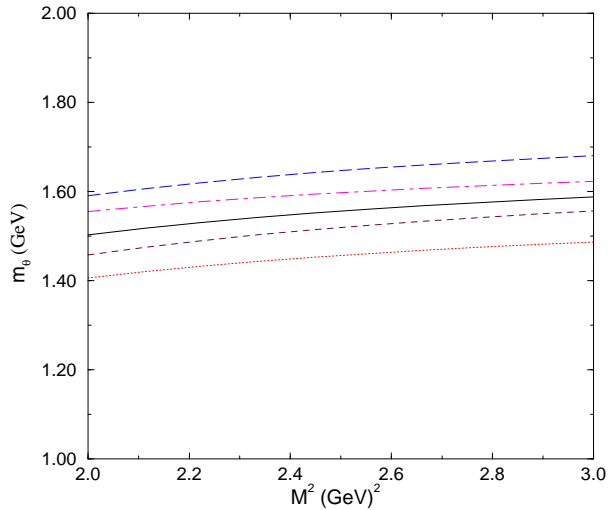,width=8cm,angle=0}}
\caption{$s_0$ and $t$ dependence of $m_\Theta$. The solid, dotted and
long-dashed lines give $m_\Theta$ for $s_{0\Theta}=4.0~\GeV^2$, 
$s_{0\Theta}=3.6~\GeV^2$ 
and $s_{0\Theta}=4.4~\GeV^2$ respectively and $t=-1$. The dot-dashed
 and dashed 
lines give $m_\Theta$ for $t=-0.8$ and $t=-1.1$ respectively and
$s_{0\Theta}=4.0~\GeV^2$.}
\end{figure}

The effect of changing the current parameter, $t$, is similar to the effect
of changing the continuum threshold. In Fig.~3 we show the mass
of $\Theta^+$ for different values of the continuum threshold and
the current parameter. Decreasing the value of $t$ (in modulus) increases
the value of both masses keeping the difference approximately constant. 
In Fig.~3 we show an example with $t=-0.8$, but $t$
can be even smaller without leading to a big change in the masses. 
For 
$t=-0.5$, for example, we still have a value that is close to the 
result for $t=-1$ and $s_{0\Theta}=4.4~\GeV^2$ (the upper curve in Fig.~3).
Increasing $t$ has a stronger effect in decreasing the masses, the effect
being even bigger for $\Theta^+$. For $t=-1.2$ for instance, the mass 
difference is only of order of $60~\MeV$ and for $t=-1.3$ it is negative.

Fixing $M^2=2.5~\GeV^2$ and considering the ranges 
$3.6~\GeV^2\leq s_{0\Theta}\leq
4.4~\GeV^2$, $3.2~\GeV^2\leq s_{0N}\leq4.0~\GeV^2$ and $-1.1\leq t\leq-0.8$
our results for both pentaquark masses are
\beq
m_\Theta=1.55\pm0.10~\GeV,\;\;\;\;\;\;\;\;m_N=1.44\pm0.11~\GeV\,,
\eeq
and $m_\Theta-m_N=110\pm10~\MeV$, in a very good agreement with
the experimental result.

In \cite{zhu} the value of $M_0=1.56 \pm 0.15 \GeV$ was found for 
the isoscalar 
pentaquark, in agreement both with experiment and with our QCDSR calculation. 
In contrast to our result, the sum rule obtained in \cite{zhu} does not depend 
on the strange quark mass $m_s$. As a consequence, the substitution 
$s \rightarrow d$ would only change the quark condensates (
$\langle\overline{s}s\rangle \rightarrow \langle\overline{q}q\rangle$) 
producing a minor change in the pentaquark masses. This suggests that 
$N^+$ and $\Theta^+$ would have a smaller mass difference than we find here.


In conclusion,
we have presented a QCD sum rule study of the $\Theta^+$ and $N^+$
pentaquarks masses using the scheme proposed by Jaffe and Wilczek, according 
to which both resonances are diquark-diquark-antiquark states. Using our 
interpolating current we were able to reproduce the experimental value of the 
$\Theta^+(1540)$ and we obtained a mass for the $N^+$ which is compatible with 
the interpretation of this state as the Roper resonance $N(1440)$.
We studied the mass
difference between these states as a function of the continuum threshold and
the current parameter, obtaining   $m_\Theta-m_N \simeq 100~\MeV$. 

The difference in the sum rules for  $\Theta^+$ and $N^+$ can be entirely 
assigned to the strange quark mass. This finding supports the analysis 
performed in \cite{jawil} and encourages us to calculate the masses of
the other antidecuplet members, in particular the masses of the states 
$\Xi^+$ and $\Xi^{--}$. The mass of these states has been estimated
to be $\sim2070~\MeV$ in ref.~\cite{dpp} and $\sim1750~\MeV$ in
ref.~\cite{jawil}. A QCDSR calculation may be useful in discriminating
between the two approaches.

\vspace{1cm}
 
\underline{Acknowledgements}: 
This work has been supported by CNPq and  FAPESP (Brazil). 


\end{document}